\documentstyle[aps,preprint,eqsecnum]{revtex}
\begin{document}
\draft
\preprint{$\hbox to 5 truecm{\hfill Alberta-Thy-32-94}\atop
{\hbox to 5 truecm{\hfill gr-qc/9411001}\atop
\hbox to 5 truecm{\hfill October 1994}}$}
\title{Action and Hamiltonian for eternal black holes}
\author{Valeri Frolov and Erik A. Martinez \footnote{electronic addresses:
frolov@phys.ualberta.ca, martinez@phys.ualberta.ca}}
\address{Theoretical Physics Institute,\\
University of Alberta,\\
Edmonton, Alberta T6G 2J1, Canada.}
\maketitle
\begin{abstract}

We present the Hamiltonian, quasilocal energy, and angular momentum for a
spacetime region spatially bounded by two timelike surfaces. The results are
appplied to the particular case of a spacetime representing an eternal black
hole. It is shown that in the case when the boundaries are located in two
different wedges of the Kruskal diagram, the Hamiltonian is of the form  $H =
H_{+} - H_{-}$,  where $H_{+}$ and $H_{-}$ are the
Hamiltonian functions for the right and left wedges respectively.
The application of the obtained results to the thermofield dynamics description
of quantum effects in black holes is briefly discussed.

\end{abstract}
\vspace{7mm}
\vfill
\eject

\section{Introduction}
\label{sec:introduction}

One of the most important problems of black hole physics concerns the
dynamical origin of black hole entropy and, more generally, the statistical
mechanical foundations of black hole thermodynamics.  A natural way to
calculate the thermodynamical parameters of a black hole is to use a
``complexification" approach \cite{GiHa1,Yo,BrMaYo}. However,  in this
framework there is simply no room for the internal degrees of freedom of the
black hole since one deals explicitly only with the Euclidean version of the
black hole exterior.
That is why it is important to begin with the usual
Lorentzian description of a black hole in order to understand the
dynamical origin of its internal degrees of freedom.

In order to describe a state of a black hole in the Lorentzian approach one
uses a spatial slice which crosses the horizon.
In the general case such a slice crosses also a singularity inside the black
hole horizon so that fixing a state requires additional data describing the
singularity.
This difficulty can be overcome by considering the `eternal version' of a black
hole, as done recently in \cite{BaFrZe}.
The main idea of this approach is as follows: a black hole which is formed as a
result of gravitational collapse is described at late time by a stationary
metric.
One can use analytical continuation to connect the late time
geometry of this configuration with the spacetime of an eternal black hole
possessing the same parameters (mass $M$, angular momentum
$J$, charge $Q$, etc.).  This analytically continued spacetime is called the
`eternal version' of the original black hole. The Penrose diagram for such a
spacetime is shown in Figure 1.
Classical or quantum excitations of the black hole at late time can be
traced back in the spacetime of its `eternal version' to some initial global
Cauchy surface.  For example, one can choose a slice $\Sigma$ representing a
spatial section at the moment of time symmetry as such a surface.  The surface
$\Sigma$ is an
Eintein-Rosen bridge with wormhole topology $R^1 \times S^2$ (Figure 2).
In the general case when the black hole is not static and possesses external as
well as internal excitations this construction allows one to relate the
excitations of a real black hole at late time with the corresponding field
perturbations at $\Sigma$. The gravitational field perturbations on $\Sigma$
can be identified with small deformations of the three-geometry of the
Einstein-Rosen bridge. This construction can be generalized to the situation
when the  perturbations are not small. In this case the initial data on
$\Sigma$ can be
described as  (not necessarily small) deformations of the Einstein-Rosen bridge
plus initial data for the non-gravitational degrees of freedom on the distorted
Einstein-Rosen bridge
provided the constraint equations are satisfied.
In the framework of this construction one may consider, instead of different
states of
a real black hole at late time, different states of the
corresponding Einstein-Rosen bridge (or the corresponding deformations of it).
The latter problem appears to be much simpler analytically. In particular, it
allows a simple description of the internal degrees of freedom of a black hole
\cite{BaFrZe}.
The bifurcation surface of the Killing horizon is a two-dimensional
surface of  minimal area on $\Sigma$ which divides the surface
$\Sigma$ in two parts denoted by $\Sigma_+$  and $\Sigma_-$. The initial data
at $\Sigma_-$ naturally define the internal degrees of freedom of a black hole.

A black hole considered as a part of a thermodynamical system can be in
equilibrium with surrounding thermal radiation provided the temperature of
the radiation (measured at infinity) coincides with the black hole temperature.
This equilibrium is stable only if the black hole is contained inside
a cavity of finite size where the boundary data of the appropriate statistical
ensemble are specified. (For example, for a non-rotating black hole
inside a spherical boundary of radius $r_0$ with fixed temperature $T_0$,
 stable equilibrium requires $r_0 \leq 3M$.)
The important role played by boundaries in formulations of black hole
thermodynamics has been stressed by York and collaborators \cite{Yo,BrMaYo}.

An interesting attempt to develop thermodynamical formulations of black holes
directly in the Lorentzian approach was made in Ref.\cite{BrYo2}. However, as
indicated above, the existence of singularities in the black hole interior
creates potential difficulties in such an approach. That is the reason why it
is interesting to develop the Lorentzian approach beginning directly with the
eternal version of a black hole \cite{FrMa}.
However, a price to be paid is that a single external two-dimensional boundary
does not restrict anymore a finite three-dimensional
volume on a spacelike slice with wormhole topology.
In order to analyze problems involving a finite volume for such
spaces one needs to consider two (holomorphic) surfaces, each one with
two-sphere
topology.
The main aim of this paper is to generalize the results of Ref. \cite{BrYo1} to
the case of eternal black holes and spacetime regions contained between two
timelike boundaries. Namely, we construct the gravitational Hamiltonian and
generalize the quasilocal energy and angular momentum definitions for this
kind of situations.
Special attention is paid to the case when
the external boundary is located, as
usual, in the black hole exterior while the other boundary (the internal one)
is located either at the external part $\Sigma_+$ or at the internal part
$\Sigma_-$ of the Einstein-Rosen bridge.

The construction of Hamiltonian and quasilocal energy for a spatially bounded
region of an eternal version of a black hole is useful when considering quantum
properties of black holes.
Recently  a no-boundary ansatz (analogous to the Hartle-Hawking
no-boundary ansatz in quantum cosmology) has been proposed to define a wave
function of a black hole \cite{BaFrZe}. Such a wave function is a functional of
the configuration space of deformations of the Einstein-Rosen bridge.
For field perturbations this wave function
describes a Hartle-Hawking state of linear fields propagating in the background
of a black hole. Restriction to a part of the system (either internal or
external) generates a thermal density matrix.
The thermal nature of the density matrix is directly related with the special
structure of the Hamiltonian $H$ for small perturbations, which is of the form
$H = H_+ - H_-$, where $H_+$ and $H_-$ are the Hamiltonians for perturbations
on  the  external  and internal parts of an Einstein-Rosen bridge respectively.
Israel \cite{Is} has noted that this situation is in complete
accord with the thermofield dynamics approach to thermodynamics developed in
Ref.~\cite{UmTa}. The results obtained in this
paper allow one in principle to develop this analogy beyond
perturbations to the strong
gravity regime, in which all the quantities, and in particular the Hamiltonian,
refer to the gravitational field  of the black hole itself.
The total Hamiltonian  for the system is encoded in the
boundary conditions and posseses  the general form $H = H_+ - H_-$, where
$H_+$
and $H_-$ are the gravitational Hamiltonians for the external and internal
regions of a distorted black hole.
A full generalization of the thermofield dynamics approach to black holes
can be related therefore with the study of geometrodynamics of an `eternal
version' of a black hole.

The paper is organized as follows.
We begin by discussing in Section II the geometry of the spacetime of a
Schwarzschild-Kruskal
eternal black hole and its natural foliation connected with the global Killing
vector field $\xi _t$ (``tilted foliation"). We describe  generalizations of
the
``tilted foliations" for a general spacetime (without a Killing vector) and
general
kinematical relations applicable to ``tilted" as well as to the standard
(``untilted") foliations. The general expressions for  Hamiltonian, quasilocal
energy and angular momentum are obtained by using a ``tilted foliation" in
Sections III, IV, and V respectively. The dynamics of the standard
(``untilted") foliation is discussed in Section VI. The possible relation of
the obtained results with the thermofield dynamics approach to black hole
thermodynamics is  discussed in Section VII.
We use the sign conventions of Ref.~\cite{MTW}, and units are chosen so that
$G=c=\hbar=1$.

\section{Foliations and Kinematics}
\label{sec:foliations}
\subsection{Topology}
\label{subsec:topology}

Before considering the more general case of a distorted black hole, we describe
some properties of stationary vacuum eternal black hole solutions. Consider at
first the case of a static, non-rotating black hole. The corresponding
Schwarzschild-Kruskal line element reads:
\begin{equation}
ds^2 = {{-32 {\cal M}^3}\over{r}} e^{-r /{2\cal M}}
\, dU \, dV + r^2 d\Omega ^2 \ , \label{Sch}
 \end{equation}
where ${\cal M}$ represents the Arnowitt-Deser-Misner (ADM)
mass of the static  black hole \cite{ADM},
$d\Omega^2 $ is the line element of the unit two-sphere, and the radial
coordinate is regarded as a function of the ingoing and outgoing null
coordinates $(U,V)$ as
 \begin{equation}
(1 - r/{2\cal M}) e^{r /{2\cal M}} = U V \ .
\end{equation}
The corresponding Penrose diagram is shown in Figure 1.  The
Schwarzschild-Kruskal spacetime \cite{MTW} is the union of four regions
(wedges)  $R_+$, $R_-$, $T_+$, and $T_-$. The regions $R_+$ and $R_-$ are
asymptotically flat . In the $R_+$  region, the Kruskal coordinates $(U,V)$
satisfy $U < 0$, $V>0$, while in $R_-$, $U > 0$, $V<0$. The Killing vector
field $\xi _t$ connected with the time symmetry is
\begin{equation}
{\xi_t }^\mu \partial
_\mu = {1 \over{{4\cal M}}} \bigg( V \partial _V - U \partial _U \bigg)\ .
\end{equation}
This Killing vector is timelike in $R_{\pm}$, spacelike in $T_\pm$, and becomes
null at the future ($H^+$) and past ($H^- $) horizons.
It is future oriented in $R_+$ and  past oriented in $R_-$, whereas it vanishes
at the bifurcation surface ${\tilde S}_0 =  H^+ \cap H^-$.

Surfaces orthogonal to $\xi_t$ are defined by the equation
$t={\rm constant}$, where the scalar function $t$ is defined as
\begin{equation}
t = 2{\cal M} \ln \, \bigg| {-V\over{U}}\bigg| \ .
\end{equation}
The line element (\ref{Sch}) is invariant with respect to the discrete
symmetries
\begin{eqnarray}
{\bf I} &:& U \to - U ;\, V \to -V  \ ,\nonumber \\
{\bf T} &:& U \to - V ;\, V \to -U  \ .
\end{eqnarray}
The surface ${\tilde \Sigma}$ is defined by the equations $t = 0$ or,
equivalently, $U+V =0$. This surface is invariant under $T$-reflections and,
(as well as any other surface $t={\rm constant}$)  has Einstein-Rosen bridge
topology $R^1 \times S^2$  (Figure 2).  We denote by  ${\tilde \Sigma}_{\pm}$
the parts of  ${\tilde \Sigma}$ lying in $R_\pm$ respectively.   The line
element (\ref{Sch}) restricted to the surface  ${\tilde \Sigma}$ reads
\begin{equation}
dl ^2 = dy^2  + r^2 (y)  d\Omega ^2 \ ,
\end{equation}
where the quantity $y$, defined by
\begin{equation}
dy = \pm {{dr}\over {\sqrt{1 - 2 {\cal M}/r}}} \ ,
\end{equation}
represents the proper geodesic distance from the ``throat" of the bridge
located
at $r = r(y=0) = r_+ = 2 {\cal M}$.
We choose $y$ to be positive in ${\tilde \Sigma}_+$ and negative  in ${\tilde
\Sigma}_-$.  The set $(t, y, \theta, \phi )$ can be used as canonical
coordinates everywhere in $R_\pm$ outside the  surface ${\tilde S}_0$. These
coordinates are right-oriented in $R_+$ and left-oriented in $R_-$.

The second homotopy group of $\Sigma \equiv R^1 \times S^2$ is
$\pi_2 (R^1 \times S^2) = {\bf Z}$. This means that,
besides the two-dimensional topologically spherical surfaces in
${\tilde \Sigma}$ which can be contracted to a point, there also exist
in  ${\tilde \Sigma}$ non-contractible two-dimensional spheres.  An example
of such a  sphere is the bifurcation surface ${\tilde S}_0$ or any other
surface obtained
from it by continuous transformations.  Therefore, in order to restrict any
finite three-dimensional volume on the $t={\rm const.}$ hypersurface it is
necessary
to have as boundaries {\it two} such topological spheres
(${{\tilde S}_+}$ and ${{\tilde S}_-}$).
In the particular case when the boundary surfaces are round spheres, they are
defined by the equations $y = y_+$ and $y = y_-$.
The spatial region restricted by these spheres has topology $I \times S^2$,
where $I$ is a finite interval.
In mathematical literature, this region is known as cobordism between the
boundary surfaces ${\tilde S}_+$ and ${\tilde S}_-$.

A natural way of describing dynamics of fields propagating on the eternal black
hole background is to use the Killing time $t$  as the evolution parameter.
Two different physical situations are possible in principle: (1) both boundary
surfaces are located in the same region of the spacetime (either $R_+$ or
$R_-$), and (2) the boundary surfaces are located in different regions. In case
(1), the $t={\rm const.}$ slices do not intersect each other, so that the
foliation is of the standard form. In case (2), two different slices ${\tilde
\Sigma}'$ ($t=t'$)
and  ${\tilde \Sigma }''$ ($t=t''$) intersect at the two-dimensional
bifurcation
sphere ${\tilde S}_0$.
We call the sequence of slices ${\tilde \Sigma}_t$ (defined by the
equation $t = {\rm constant}$, with $t' \leq t \leq t''$) which intersect at
the same bifurcation surface ${\tilde S}_0$ the ``tilted foliation".
For a ``tilted foliation" the spacetime domain $M$ lying between
${\tilde \Sigma}'$ and
${\tilde \Sigma} ''$  consists of two wedges  $M_+$ and $M_-$ located in the
right  ($R_+$) and left ($R_-$) sectors of the Kruskal diagram.
In both cases, if we restrict ourselves by considering the field dynamics in a
finite region of spacetime restricted by boundary slices ${\tilde \Sigma}'$ and
${\tilde \Sigma }''$, we need also to specify boundary conditions at two
timelike
boundaries, which we denote by ${\tilde{{B_{\scriptscriptstyle {\pm}}}}}$.
Each of these boundaries is the result of time evolution from $t'$ to $t''$ of
a
topological sphere homotopic to ${\tilde S}_0$.
Both type of foliations mentioned above can be defined in the more general case
when one does not restrict oneself by considering the dynamics with respect to
the Killing time $t$.

We describe at first the construction of the ``tilted" and ``untilted"
foliations for the Kerr-Newman spacetime.
In the case of a rotating black hole, the Killing vector $\xi_t$
(uniquely defined as the Killing vector which is  timelike at spatial infinity)
is not hypersurface orthogonal anymore. Nevertheless, its linear combination
with the Killing vector $\xi_\phi$ (corresponding to axial symmetry)
\begin{equation}
u^\mu = A\, ( \xi_t ^\mu + \omega \,  \xi_\phi ^\mu )
\end{equation}
defines a unit vector $u^\mu$  normal to a hypersurface $\Sigma _t$, provided
that
\begin{equation}
\omega = -  {{g_{t\phi}}\over{g_{\phi \phi}}} ;\, \,
\,A = \sqrt{
{
{g_{\phi \phi}}
 \over
{ g_{t\phi}^2 - g_{tt} g_{\phi \phi}}
 }}\ .
 \end{equation}
Here  $g_{\mu\nu}$ denotes the coefficients of the Kerr-Newman
metric in the standard Boyer-Lindquist coordinates \cite{MTW}.
The hypersurface $\Sigma_t$ is defined by the equation $t = {\rm constant}$.
It is easy to verify that  $\xi_\phi  \cdot u = 0 $, so that  the
$u$-observers are zero angular momentum observers \cite{Ba}.

The topology of the surface $t = 0$ (as well as any other  surface $t = {\rm
constant}$) in the Kerr-Newman spacetime  is the same as the topology of the
$t=0$ slice for a Schwarzschild-Kruskal black hole, namely, $R^1\times S^2$.
The $t = {\rm constant}$ slices cross each other at a bifurcation
surface ${\tilde S}_0$ with the topology of a two-sphere.
The hypersurfaces $\Sigma_t$ form a``tilted foliation" for the Kerr-Newman
spacetime. The ``untilted" foliation  is generalized in a similar way.

The constructions of both ``tilted" and ``untilted" foliations can be easily
generalized to the case of a distorted black  hole.  Let us consider
a three-dimensional surface, denoted by $\Sigma$, with Eintein-Rosen bridge
topology ($R^1 \times S^2$) and whose three-geometry is such that the
corresponding (distorted) Einstein-Rosen bridge connects two asymptotically
flat spaces.
We choose this three-geometry and its time derivatives in such a way that the
gravitational constraint equations are satisfied. These initial data can be
used  to determine the subsequent spacetime evolution by using the vacuum
Einstein equations.
We assume that this evolution defines a regular spacetime in some domain $M$
lying both to the
future and to the past of the initial Cauchy surface $\Sigma$.  We also assume
that it is possible to choose two different spacelike Cauchy surfaces
$\Sigma '$ and $\Sigma ''$ in this domain which intersect each other at the
two-dimensional, topologically spherical spacelike surface $S_0$.
Consider a differential one-to-one map $\Psi$  of the domain $M$ into  the
Schwarzschild-Kruskal spacetime possessing  the folowing properties:
\begin{equation}
\Psi (\Sigma ') =  {\tilde \Sigma }'  ;\, \,  \Psi (\Sigma '') = {\tilde \Sigma
}''  ;\, \, \Psi (B_\pm) = {\tilde B_\pm} ; \, \, \Psi (S_0) = {\tilde S}_0 ;
\,
\, \Psi (S_\pm) = {\tilde S_\pm}  \  .
\end{equation}
This map transforms the region $A$ lying between $\Sigma '$ and $\Sigma ''$
into the region ${\tilde A}$ lying between ${\tilde \Sigma} '$ and
${\tilde \Sigma} ''$.
We can use this map to provide foliations for the distorted black hole
spacetime.
These foliations are obtained by simply transfering the foliations of
the Schwarzschild-Kruskal spacetime by the map $\Psi$ onto $M$. Certainly there
exist an infinite set of maps $\Psi$ that accomplish this, and one can choose
anyone of them.
For a chosen map $\Psi$ the canonical coordinates $(t, y,\theta, \phi)$
introduced above for the ``tilted foliation" of the Schwarzschild-Kruskal
spacetime can be transfered from ${\tilde A}$ to the spacetime region $A$ on
$M$.
The coordinates $(t, y, \theta, \phi)$  are right-oriented in $R_+$ and left
oriented in $R_-$.

\subsection{Kinematics}
\label{subsec:kinematics}

We discuss now the kinematics of a  general spacetime $M$
whose line element has the form
\begin{equation}
 ds^2 = -N^2 dt^2 + h_{ij} (dx^i + V^i dt)(dx^j + V^j dt)\ ,
\end{equation}
where $t$ is the time coordinate connected with the chosen foliation, and $N$
the  corresponding lapse function. The four-velocity is defined by
$u_\mu = -N \, \partial_\mu t$,
and the lapse function $N$ is defined so that  $u \cdot u = -1$.
The four-velocity $u^\mu$ is the timelike unit vector  normal to the
spacelike hypersurfaces ($t = {\rm constant}$) denoted by $\Sigma_{t}$.
The  vector $t^\mu$ that connects points with the same spatial coordinates in
different slices is  related to the four-velocity $u^\mu$  by
\begin{equation}
t^\mu = Nu^\mu + V^\mu \ ,\label{tvector}
\end{equation}
so that
$V^{i}= h^i _0 = - N u^i$ is the shift vector.
We choose the spacelike boundaries to coincide with surfaces of fixed values of
$t$, so that the time coordinate $t$ is the scalar function that uniquely
labels
the foliation. We assume that the surfaces
$\Sigma ' \equiv \Sigma_{t'}$, and   $\Sigma '' \equiv \Sigma_{t''}$
are the spacelike boundaries of the spacetime region under consideration.

We assume that a three-dimensional timelike boundary $B$ consists of two
disconnected parts ${B_{\scriptscriptstyle +}}$ and ${B_{\scriptscriptstyle
-}}$: $B = {B_{\scriptscriptstyle +}} \cup {B_{\scriptscriptstyle -}}$. The
spacelike normal $n^{i}$ to the three-dimensional boundaries $B$ is  defined to
be outward pointing at the boundary ${B_{\scriptscriptstyle +}}$, inward
pointing at the  boundary ${B_{\scriptscriptstyle -}}$, and  normalized so that
$n \cdot n=+1$.   In what follows we also assume that  the foliation  is
restricted by the conditions $(u \cdot n)|_{B} = 0$.

In parallel to  Ref.~\cite{BrYo1}, the metric and extrinsic curvature of
$\Sigma_t$ as a surface embedded in $M$ are denoted $h_{ij}$ and
$K_{ij} = -h_i ^{\ k} \nabla_{k}u_{j}$ respectively, while the metric and
extrinsic curvature of the boundaries $B$ as  surfaces embedded in $M$ are
$\gamma_{ij}$ and  $\Theta_{ij} = -\gamma_i ^{\ k} \nabla_{k} n_{j}$.
We denote by $h$ and $\gamma$ the determinants of the three-dimensional metrics
$h_{ij}$ and $\gamma_{ij}$.
The intersection of the boundaries ${B_{\scriptscriptstyle +}}$ and
${B_{\scriptscriptstyle -}}$
with  $\Sigma_t$ are
(topologically) spherical two-dimensional surfaces
denoted by ${{S_{\scriptscriptstyle +}}}$ and ${{S_{\scriptscriptstyle -}}}$
respectively.
The induced metric and extrinsic curvature of the boundaries $S$ as surfaces
embedded on $\Sigma$ are  ${\sigma}_{ab}$ and  $k_{ab} = -\sigma_{a}^{\ k}\,
D_{k} n_{b}$ respectively.  The normal vector  $n^{i}$ to ${B}$ is also the
normal vector to ${S}$. The simbols $\nabla$ and $D$ denote covariant
differentiation with respect to the metrics $g_{{\mu\nu}}$ and  $h_{ij}$
respectively. The metric tensors for the different surfaces can be written as
\begin{eqnarray}
h_{\mu\nu} &=& g_{\mu\nu} + u_\mu u_\nu \ , \nonumber \\
\gamma_{\mu\nu} &=& g_{\mu\nu} - n_\mu n_\nu \ ,\nonumber \\
\sigma _{\mu\nu} &=& g_{\mu\nu} + u_\mu u_\nu - n_\mu n_\nu \ . \label{metrics}
\end{eqnarray}
In particular, we have
\begin{eqnarray}
\sqrt{-g} &=& |N| \sqrt{h}\ , \nonumber \\
\sqrt{-{\gamma}} &=& |N| \sqrt{\sigma} \ , \label{det}
\end{eqnarray}
where $\sigma$ is the determinant of the two-dimensional metric $\sigma_{ab}$.

As shown in \cite{BrYo1}, the extrinsic curvature tensor $\Theta^{\mu\nu}$ can
be written as
\begin{equation}
\Theta^{{\mu\nu}} = k^{{\mu\nu}} + u^\mu u^\nu n_\alpha a^\alpha + 2
\sigma^{\alpha ( \mu} u^{\nu )} n^\beta K_{\alpha \beta}\ . \label{Thetamunu}
\end{equation}
The traces of $\Theta_{{\mu\nu}}$ and $k_{{\mu\nu}}$ are related by
\begin{equation}
\Theta = k - n_\beta a^\beta \ ,\label{traces}
\end{equation}
where  $a^\mu =  u^\alpha \nabla _{\alpha} u^\mu $ is the acceleration of the
timelike normal $u^{\mu}$.

For the usual (``untilted") foliation the lapse function $N$ can be chosen
everywhere positive, and there is no intersection between slices corresponding
to different time parameter. In the case of a ``tilted foliation", slices
corresponding to different time parameter $t$ intersect at the two-dimensional
bifurcation surface where the lapse function $N$ vanishes. We assume the
parameter $N$ to be positive at $\Sigma_+$ and negative at $\Sigma_-$.

\section{dynamics of a tilted foliation}
\label{sec:dynamics of a tilted foliation}

It is instructive to study the dynamics of ``tilted" and ``untilted"
foliations independently. Consider at first the ``tilted foliation" shown in
Figure 3.  The region
lying between the two spacelike Cauchy surfaces  $\Sigma '$ and $\Sigma ''$
consists of two wedges $M_+$ and $M_-$  intersecting at a two-dimensional
surface $S_0$.  The symbol $\Sigma_{t\,(\pm)}$ denotes the part of $\Sigma_{t}$
located in $M_{\pm}$. We also denote those parts of  $\Sigma '$ and $\Sigma ''$
which are the spacelike boundaries of the wedges  $M_{\pm}$ as $\Sigma_{\pm} '$
and  $\Sigma_{\pm} '' $.
The  lapse function $N$ is positive (negative) at $M_+$ ($M_-$) and equals
zero at the bifurcation surface. The vector $u^\mu$ is
future oriented in $M_+$ and past oriented in  $M_-$.

The covariant form of the gravitational action for this foliation with  fixed
three-geometry at the boundaries of $M$ is
\begin{eqnarray}
S &=&
{1\over2\kappa} \int_{M_+} d^4 x \,\sqrt{-g}\,\Re  + {1\over\kappa}
\int_{({\scriptscriptstyle +})t'}^{t''} d^3 x \,\sqrt{h}\, K
 - {1\over\kappa} \int_{{B_{\scriptscriptstyle +}}} d^3x \,
\sqrt{-\gamma}\, \Theta \nonumber \\
&-&{1\over2\kappa} \int_{M_-} d^4 x \,\sqrt{-g}\,\Re
+ {1\over\kappa} \int_{({\scriptscriptstyle -})t'} ^{t''} d^3 x \,\sqrt{h}\, K
- {1\over\kappa} \int_{{B_{\scriptscriptstyle -}}} d^3 x \,
\sqrt{-\gamma}\, \Theta \ ,\label{bigaction}
\end{eqnarray}
where  $\Re$ denotes the four-dimensional scalar curvature
and $\kappa \equiv 8 \pi$.  It is assumed that
the integrations are taken over the coordinates $x^\mu$ which have the same
orientation as the canonical coordinates $(t, y, \theta, \phi)$ of the ``tilted
foliation". The negative sign for the integration over $M_-$ reflects the fact
that the canonical coordinates are left oriented in this region.  Besides the
volume term, the action $S$ contains also boundary terms. The notation
$\int_{({\scriptscriptstyle \pm})t'}^{t''}$  represents an  integral over the
three-boundary $\Sigma_{\pm}$ at $t''$ minus an integral over the
three-boundary
$\Sigma_{\pm}$ at $t'$.
The action (\ref{bigaction}) has to be complemented \cite{GiHa1,BrYo1,BrHa}
with a subtraction term $S^0$ (dependent only on the
boundary intrinsic geometry) which will be discussed later.

We write the Hamiltonian form of the action following the standard
procedure by recognizing that there exists a direction of time at the
boundaries inherited by the time vector field $t^\mu $ defined at  $B$.
Under a $3+1$ spacetime split, the four-dimensional scalar curvature is
\begin{equation}
\Re = R + K_{\mu\nu} K^{\mu\nu} - (K)^2 -2\nabla_\mu (K u^\mu + a^\mu)\ ,
\label{Re}
\end{equation}
where $R$ denotes the scalar curvature of the three-dimensional spacelike
hypersurface $\Sigma$. By the use of Gauss' theorem, the conditions
\begin{equation}
u \cdot n |_{B} = 0,\,  u \cdot a = 0,\,  u \cdot u = -1,\,n \cdot n = 1 \ ,
\end{equation}
and Eqns. (\ref{metrics})-(\ref{traces}) and (\ref{Re}), one can rewrite the
total action in the form
\begin{eqnarray}
S &=& {1\over2\kappa} \int_{M_+} d^4x  \sqrt{-g}\,
\bigl[R + K_{\mu\nu} K^{\mu\nu} - (K)^2 \bigr]
-{1\over\kappa} \int_{{B_{\scriptscriptstyle +}}} d^3x \sqrt{- \gamma}\, k
\nonumber \\
&-&{1\over2\kappa} \int_{M_-} d^4x  \sqrt{-g} \,
\bigl[R + K_{\mu\nu} K^{\mu\nu} - (K)^2 \bigr]
-{1\over\kappa} \int_{{B_{\scriptscriptstyle -}}} d^3x \sqrt{-\gamma}\, k \ ,
\label{littleaction}
\end{eqnarray}
where $k = - D_{\mu}n^{\mu}$ on ${B_{\scriptscriptstyle -}}$.
The regions $M_+$ and $M_-$ intersect at the cusp-like part $S_0$ of
their boundaries.  Generically  such a cusp gives an additional
contribution to the action \cite{BrHa}.  For the ``tilted foliation" the cusp
contributions  connected with the regions $M_+$ and $M_-$ have opposite signs.
The regularity of the spacetime in the vicinity of $S_0$ implies that their
absolute values are the same. That is why the action (\ref{littleaction}) does
not contain any cusp contributions connected with $S_0$.

The momentum $P^{ij}$ conjugate to the three-metric $h_{ij}$ of $\Sigma$ is
defined  as
\begin{equation}
P^{ij} = {1\over2\kappa} \sqrt{h}\, (K h^{ij} - K^{ij}) \ , \label{Pij}
\end{equation}
and the definition of $K_{\mu \nu}$ implies
\begin{equation}
K_{ij} = -\nabla _{i} u_j =
-{1\over {2N}}\, \Bigl[ \dot h_{ij} - 2D_{(i} V_{j)}
\Bigr]\ , \label{Kij}
\end{equation}
where the dot denotes differentiation with respect to the  time $t$.

Observe that  by virtue of (\ref{det}), the kinetic
part of the volume integral in (\ref{littleaction}) for both
boundaries  ${B_{\scriptscriptstyle +}}$ and ${B_{\scriptscriptstyle -}}$ can
be
written in the desired  form
\begin{equation}
 2\kappa \, \Bigl[ P^{ij}
   \dot h_{ij} -  2P^{ij} D_i V_j - 2\kappa N G_{ijk\ell} P^{ij}
   P^{k\ell}\Bigr] \ ,\label{kinetic}
\end{equation}
where $G_{ijk\ell} = (h_{ik} h_{j\ell} + h_{i\ell} h_{jk} - h_{ij} h_{k\ell})
/(2\sqrt{h}) $.
The second term in the right hand side of (\ref{kinetic}) can be also
rewritten since $ -2 P^{ij} D_i V_j = -2[ D_i(P^{ij} V_j) - V_j D_i P^{ij}]$
and
\begin{equation}
-2\int d^4 x  D_i(P^{ij} V_j) =
-2\int_{{B_{\scriptscriptstyle +}}} d^3 x\, n_i \Big( P^{ij} V_j/\sqrt{h}
\Big) \sqrt{\sigma}
+2\int_{{B_{\scriptscriptstyle -}}} d^3 x \, n_i \Big( P^{ij} V_j/\sqrt{h}
\Big) \sqrt{\sigma} \ ,
\end{equation}
with the cusp contributions at $S_0$ cancelling identically.
Substitution of (\ref{kinetic}) into (\ref{littleaction}) gives us the desired
Hamiltonian form of the action. In terms of the quantities \cite{BrMaYo,BrYo1}
\begin{equation}
\varepsilon \equiv  \bigl( k/\kappa\bigr) \label{edensity} \ ,\,
        j^i \equiv -2  \sigma^i_{\ k} n_{\ell} P^{ k\ell}/\sqrt{h}\ ,
\label{mdensity}
\end{equation}
the action finally becomes
\begin{equation}
 S = \int_M d^4x  P^{ij} \dot h_{ij} - \int H dt \ ,\label{Hamaction}
\end{equation}
with the gravitational Hamiltonian $H$ given by
\begin{equation}
H =\int_{\Sigma} d^3x \bigl(N{\cal H} + V^i {{\cal H}_i} \bigr)
+ \int_{{S_{\scriptscriptstyle +}}} d^2x \sqrt{\sigma} \bigl(
N \varepsilon - V^i  j_i\bigr)
 - \int_{{S_{\scriptscriptstyle -}}} d^2x \sqrt{\sigma} \bigl
( N \varepsilon - V^i  j_i\bigr)\ .
\label{Ham}
\end{equation}
The Hamiltonian posseses both volume and boundary contributions. The volume
part involves the gravitational contribution to the Hamiltonian and momentum
constraints:
\begin{eqnarray}
{\cal H} &=& (2\kappa) G_{ijk\ell}\, P^{ij}\, P^{k\ell} - \sqrt{h}
      \, R/(2\kappa) \ , \nonumber \\
{{\cal H}_i} &=& -2 D_j \, P_i^{\ j} \ . \label{constraints}
\end{eqnarray}
Notice that for stationary solutions, the first term of (\ref{Hamaction}) is
zero. If one imposes the gravitational constraint equations, the volume part of
the Hamiltonian (\ref{Ham}) is zero.
The  Hamiltonian  can be written in the suggestive form
\begin{equation}
H =H_{+} - H_{-} \ , \label{Hamiltonian}
\end{equation}
where the Hamiltonian $H_+$ of $M_+$ and the Hamiltonian $H_-$ of $M_-$ are
\begin{eqnarray}
H_{+} &=&\int_{{S_{\scriptscriptstyle +}}} d^2x \sqrt{\sigma} \bigl(
N \varepsilon - V^i  j_i \bigr)
\nonumber \\
H_{-} &=& \int_{{S_{\scriptscriptstyle -}}} d^2x \sqrt{\sigma} \bigl
( N \varepsilon - V^i  j_i \bigr)
 \ .\label{H+H-}
\end{eqnarray}

\section{quasilocal energy}
\label{sec:qle}

The quasilocal energy can be defined, following Ref.\cite{BrYo1}, as the value
of the Hamiltonian that generates unit time translations orthogonal to the
two-dimensional spacelike boundaries. The  total quasilocal energy is
therefore the value of the Hamiltonian (\ref{Hamiltonian}) (with $V_i = 0$)
such that $|N| = 1$ at both ${S_{\scriptscriptstyle +}}$ and
${S_{\scriptscriptstyle -}}$, namely
\begin{equation}
E_{\rm\scriptscriptstyle tot} = E_{+} - E_{-}\ , \label{energy1}
\end{equation}
with
\begin{eqnarray}
E_{+}&=&\int_{{S_{\scriptscriptstyle +}}} d^2x \sqrt{\sigma} (\varepsilon -
{\varepsilon}^0 )\ ,
\nonumber \\
E_{-}&=& - \int_{S_{\scriptscriptstyle -}} d^2x \sqrt{\sigma}
(\varepsilon - {\varepsilon}^0)\ .
\label{energy2}
\end{eqnarray}
This is the quasilocal energy of a spacelike hypersurface
$\Sigma = \Sigma _+ \cup \Sigma _-$ bounded by two spacelike boundaries
${S_{\scriptscriptstyle +}}$
and ${S_{\scriptscriptstyle -}}$ located in the two regions ${M_+}$ and ${M_-}$
respectively.
The quantity $\varepsilon$, defined in terms of the trace $k$ of  extrinsic
curvature in (\ref{edensity}), is seen to be proportional to the surface energy
density of the gravitational field. It is important to stress that for both
boundaries $k=-D^{\mu}n_{\mu}$.
The signs of $E_+$ and $E_-$ are directly connected with the signs of extrinsic
 curvatures of the surfaces $S_+$ and $S_-$ for the chosen orientation of the
normal vector $n_{\mu}$. In the particular case of an eternal black hole, both
the energies $E_+$ and $E_-$ are positive provided the surface $S_+$ is chosen
in the right wedge $M_+$ and the surface $S_-$ is chosen in the left wedge
$M_-$. The expressions (\ref{Hamiltonian})-(\ref{H+H-}) for the Hamiltonian as
well as expressions (\ref{energy1})-(\ref{energy2}) for the quasilocal energy
are valid for any ``tilted foliation".

We have included in these expression the relevant subtraction terms
$\varepsilon ^0$ for the energy \cite{GiHa1,BrYo1}. The role of the subtraction
term is to normalize the action with the given boundary conditions to zero for
a
reference spacetime (usually taken to be flat spacetime). The reference space
is a fixed hypersurface of this spacetime and $k^0$ the trace of extrinsic
curvature in this space of a two-dimensional surface whose induced metric is
${\sigma}_{ab}$.
\footnote{The  evaluation of this  term is delicate, even for a single boundary
surface, since there is no definite prescription of how to to embed uniquely an
arbitrary surface whose intrinsic metric is ${\sigma}_{ab}$ into a reference
space.}
As indicated in Eqn.(\ref{energy2}) one needs to make the corresponding
subtraction for each of the boundaries $B_+$ and $B_-$ independently. As the
result of this procedure, in the limiting case when $S_\pm$ tends to asymptotic
spatial infinity the limiting values of either of the quantities
$E_+$ and $E_-$coincide with the ADM energy ${\cal M}$.

To illustrate this subtraction procedure and to gain some intuition about the
relative signs of $E_+$ and $E_-$, consider the following example, consisting
of
a static Einstein-Rosen bridge whose line element is:
\begin{equation}
ds^2 = -N^2 dt^2 +  dy^2 + r^2 (y) d\Omega ^2 \ ,
\end{equation}
where $N$ and $r$ are functions of the  proper distance coordinate
$y$ continuously defined on $M$. The boundaries ${S_{\scriptscriptstyle +}}$
and  ${S_{\scriptscriptstyle -}}$ are located at
coordinate values $y=y_+$ and $y=y_-$ respectively. The normal to the
boundaries
is $n^ \mu = \delta ^\mu _y $. It is easy to see, by using the expression
\begin{equation}
k = - {1 \over {\sqrt{h}}} \, \big( \sqrt{h}\, n^\mu \big)_{,\mu}
\end{equation}
that the trace $k$  at either boundary  is
\begin{equation}
k = {{-2 r_{,y}}\over{r}}\ . \label{explicitk}
\end{equation}
The value of $k$ depends on the function $r_{,y}$, which is positive for
${S_{\scriptscriptstyle +}}$
and negative for ${S_{\scriptscriptstyle -}}$.
The trace $k$ associated with the subtraction term is taken to be
$k^0 = -2/r$ for ${B_{\scriptscriptstyle +}}$ and $k^0 = 2/r$ for
${B_{\scriptscriptstyle -}}$.  This implies that the quasilocal energy (with
subtraction terms included) can be written as
\begin{eqnarray}
E &=& E_+ - E_-  \nonumber \\
  &=& \bigg(r  \Big[ 1 - \big|r_{,y}\big| \Big] \bigg)_{y = y_+} -
      \bigg(r  \Big[ 1 - \big|r_{,y}\big| \Big] \bigg)_{y = y_-} \ .
\label{explicitE}
\end{eqnarray}
The energies  $E_+$ and $E_-$ have the same relative sign. The total energy is
zero for  boundary conditions symmetric with respect to the bifurcation surface
$S_0$. The energy  $E_+ $($E_-$) tends  to the ADM mass $\cal M$ whenever the
boundary ${B_{\scriptscriptstyle +}}$ (${B_{\scriptscriptstyle -}}$) tends to
right  (left) spatial  infinity and the gravitational constraints are inforced.
As an illustration of Eqn. (\ref{explicitE}), consider the case when the
boundary ${B_{\scriptscriptstyle +}}$ is located at right-hand infinity ($y_+ =
\infty$) and the boundary ${B_{\scriptscriptstyle -}}$ is located at $y_-$.
The energy for this configuration is
\begin{equation}
 E = {\cal M} - r \Bigg[ 1 - \bigg(1 - {{2 {\cal
M}}\over r} \bigg) ^{1/2}\, \Bigg] \ ,
\end{equation}
where $r=r({y_-})$. The energy is an even function of $y_-$ (Figure 5). It is
zero when  $|y_-| = \infty$, and reaches a minimum value  ($-{\cal M})$ at
the bifurcation surface.  The quasilocal energy  is seen to be non-positive for
all values of $y_-$ when ${B_{\scriptscriptstyle +}}$ is at right-hand
infinity.
This behavior is due to the fact that the quasilocal energy defined by
(\ref{energy1})-(\ref{energy2}) takes explicit account of gravitational
binding.

The previous example illustrates a general property of the quasilocal energy
(\ref{energy1}): for a spacetime with a bifurcation surface, the  quantities
$E_+$ and $E_-$ have the same relative sign, while the total energy is given by
the subtraction $E_+ - E_-$ \cite{FrMa}.
The same property holds  for  the Hamiltonian. This is physically very
attractive: the regions $M_+$ and $M_-$ of an eternal black hole are
symmetric to each other, the time along the respective boundaries being
determined by the global time $t$. The energies associated to
${B_{\scriptscriptstyle +}}$ and ${B_{\scriptscriptstyle -}}$ are symmetric and
the total energy  reflects the orientation reversal of the two boundaries.
The dynamical aspects of the gravitational theory for spherically symmetric
eternal black holes was recently considered by Kucha\v{r}~\cite{Ku1}. Our
expressions  (\ref{H+H-}) reproduce  his results when the black hole is
spherically symmetric and both boundaries ${B_{\scriptscriptstyle +}}$ and
 ${B_{\scriptscriptstyle -}}$
are taken to their corresponding spatial infinities.

The time direction induced at the boundaries ${B_{\scriptscriptstyle +}}$ and
${B_{\scriptscriptstyle -}}$ considered above
was inherited from the  vector $t^\mu$. While the total action is independent
of  time orientations at the boundaries, the  Hamiltonian does indeed depend on
this orientation.

\section{angular momentum}
\label{sec:angularmomentum}

The Hamiltonian form of the action given by (\ref{Hamaction}) can be used to
define charge-like quantities \cite{BrYo1}.
Consider in particular the definition  of angular momentum. The value
of the Hamiltonian function  whenever $N=0$ at the boundaries
${S_{\scriptscriptstyle +}}$ and
${S_{\scriptscriptstyle -}}$ is  $H = H_+ - H_-$ with
\begin{eqnarray}
H_{+} &=&-
\int_{{S_{\scriptscriptstyle +}}} d^2x \sqrt{\sigma} \bigl( V^i  j_i\bigr)
\nonumber \\
H_{-} &=&- \int_{{S_{\scriptscriptstyle -}}} d^2x \sqrt{\sigma} \bigl(
V^i  j_i\bigr)\ .
\end{eqnarray}
The momentum density $j^i$ of a spacelike hypersurface $\Sigma$, defined by
Eqn. (\ref{mdensity}),  is  the value per unit area of the Hamiltonian that
generates spatial diffeomorphisms in the $\partial /{\partial x^i}$ direction
on
the two-dimensional boundaries $S$.
Assume that the boundary three-metrics admit a rotational
symmetry represented by a Killing vector $\xi _\phi$.  If $\xi _\phi$ is
tangent
to both ${S_{\scriptscriptstyle +}}$ and ${S_{\scriptscriptstyle -}}$
(that is, both two-boundaries contain the orbits of $\xi _\phi$), the total
angular momentum is defined as (minus) the value of the Hamiltonian that
generates a unit rotation along  $\xi _\phi$. That is, the total angular
momentum is the value of $H$ with $N=0$ and $V^i = {\xi _\phi}^i$ at the
boundaries, namely  \begin{equation}
J_{\rm\scriptscriptstyle tot} = J_{+} - J_{-} \ ,\label{angmom1}
\end{equation}
with
\begin{eqnarray}
J_{+} &=& \int_{{S_{\scriptscriptstyle +}}} d^2x \sqrt{\sigma}\bigl(j_i
{\xi _\phi}^i \bigr)\ ,
\nonumber \\
J_{-} &=& \int_{{S_{\scriptscriptstyle -}}} d^2x \sqrt{\sigma} \bigl(j_i
{\xi _\phi}^i \bigr) \ .
\label{angmom2}
\end{eqnarray}
As for the quasilocal energy, the  angular momentum for the complete spacetime
can be expressed as the subtraction of two terms of the same functional form.
Each term is an integral over a two-surface (with unit normal orthogonal to
$\xi_\phi$) of the momentum density in the $\xi _\phi$ direction.  Notice in
particular that for symmetric boundary conditions the total angular momentum is
zero.

\section{Dynamics of untilted foliations}
\label{sec:untilted}

Consider the case of an ``untilted" foliation illustrated in Figure 4.
This foliation is regular everywhere in a region of $M$ lying
between the initial $\Sigma'$ and final $\Sigma''$ slices. The
spacetime is spatially bounded by the timelike boundaries $B_-$ and $B_+$. Let
$t$ denote the global time parameter labeling the slices $\Sigma$.
The coordinates $(y,\theta, \phi)$ are defined as in previous sections.
The foliation is such that $u ^\mu =-N {\delta^{\mu}}_{t}$, with the lapse
function defined everywhere positive. For simplicity, we consider only the
special case
$V^i =0$. The gravitational action when the
three-geometry is fixed  at the boundaries takes the form:
\begin{eqnarray}
S ={1\over2\kappa} \int_M d^4x \,\sqrt{-g} \,\Re
&+& {1\over\kappa}\int_{({\scriptscriptstyle +})t'}^{t''} d^3x \,\sqrt{h}\,K
- {1\over\kappa} \int_{{B_{\scriptscriptstyle +}}} d^3x
\,\sqrt{-\gamma}\,\Theta \nonumber \\
 &-& {1\over\kappa}
\int_{({\scriptscriptstyle -})t'}^{t''} d^3x\,\sqrt{h}\, K +
{1\over\kappa}\int_{{B_{\scriptscriptstyle -}}} d^3x \,\sqrt{-\gamma} \,\Theta
- S^0 \ ,
\end{eqnarray}
where $S^0$ represents the subtraction action.
By using the general decomposition (\ref{Kij}), the action in
Hamiltonian form becomes
\begin{equation}
S = \int_M d^4x \bigl[ P^{ij} \dot h_{ij} - N{\cal H} \bigr]
   - \int_{B_+} d^3x \sqrt{\sigma}\, N\varepsilon
   + \int_{B_-} d^3x \sqrt{\sigma}\, N\varepsilon  - S^0\ .\label{unaction}
\end{equation}
The boundary Hamiltonian is now
\begin{equation}
H =  \int_{S_+} d^2x \sqrt{\sigma}\, N( \varepsilon - \varepsilon ^0) -
     \int_{S_-} d^2x \sqrt{\sigma}\, N( \varepsilon - \varepsilon ^0)
\label{unHam}
\end{equation}
and the quasilocal energy is
\begin{eqnarray}
E_{\rm\scriptscriptstyle tot} &=& E_{+} - E_{-}\ , \nonumber  \\
E_{+}&=&\int_{S_+} d^2x \sqrt{\sigma} \,(\varepsilon - {\varepsilon}^0 ) \ ,
\nonumber \\
E_{-}&=&  \int_{S_-} d^2x \sqrt{\sigma}\, (\varepsilon -
{\varepsilon}^0)\ .  \label{une}
\end{eqnarray}
The expression (\ref{unHam}) for the Hamiltonian as
well as the expressions (\ref{une}) for the quasilocal
energy are valid for any ``untilted" foliation.
The signs of $E_+$ and $E_-$ are directly connected with the signs of extrinsic
curvatures of the surfaces $S_+$ and $S_-$ for the chosen direction of the
normal $n^{\mu}$.  Expressions (\ref{une})
are to be contrasted with Eqns.(\ref{energy1})-(\ref{energy2}) for a ``tilted
foliation".

The same results can be obtained from a covariant analysis.
Consider for simplicity a spherically symmetric spacetime.
The normal vector $n$, defined at the boundary, can be continued to the
interior of $M$ by assuming that $n \cdot \nabla y$ is positive throughout.
The scalar curvature can be written as \cite{Yo}
\begin{equation}
\Re = {{-2} \over {\sqrt{-g}}}\, \bigg( \sqrt{-\gamma}\, n^\alpha \,a_{\alpha}
\bigg)_{,y} - 2 {G^0}_0 \ .\label{sccovariant}
\end{equation}
The  action is easily obtained by applying Gauss' theorem to the first term
of (\ref{sccovariant}).
By enforcing the Hamiltonian constraint ${G^0}_0 =0$, and using Eqn.
(\ref{traces}) for the trace of extrinsic curvature, the
action reduces to the boundary action:
\begin{equation}
S_B = - \int_{{B_{\scriptscriptstyle +}}} d^3x \sqrt{\sigma}\, N\varepsilon
+ \int_{{B_{\scriptscriptstyle -}}} d^3x \sqrt{\sigma}\,  N\varepsilon
- S^0 \ ,
\end{equation}
which agrees with the action (\ref{unaction})  when the gravitational
constraints and the  staticity conditions are imposed.

Consider as a particular example a static, spherically symmetric eternal black
hole generated by an untilted foliation when both boundaries $B_+$ and $B_-$
are located in the `right' wedge $M_+$. For the natural choice of global time
$t$ as the parameter numerating the slices, one obtains the quasilocal energy
\begin{equation}
E = \bigg(r \, \Big[ 1 - |r_{,y}| \Big] \bigg)_{y = y_2} -
    \bigg(r \, \Big[ 1 - |r_{,y}| \Big] \bigg) _{y = y_1} \ .
\end{equation}
where $y=y_\pm$ denotes the location of the boundaries $B_\pm$ respectively.
Both energies $E_+$ and $E_-$ are positive when the surfaces $S_+$ and $S_-$
are in the `right' wedge $M_+$.
It can be seen that this energy is of the same form as (\ref{explicitE}).
This means that the quasilocal energy expressions for an eternal black hole
are of the same form whether the `left' boundary is located
in $M_+$ or  in $M_-$, provided the global time $t$ is chosen as the natural
scalar function labeling the foliation (that is, provided the foliation is
``untilted" when the `left' boundary lies in $M_+$ and ``tilted" when the
`left' boundary lies in $M_-$).

The expressions (\ref{une}) for the quasilocal energy of an
``untilted foliation"  formally coincide with the quasilocal energy
(\ref{energy1})-(\ref{energy2})
for ``tilted foliations" if one formally changes the direction of time $t$ at
one of the boundaries (that is, if one formally changes the direction of time
in
${B_{\scriptscriptstyle -}}$ to be $t_{-}= - t_{+}$, where $t = t_{+}$ denotes
the time direction in ${B_{\scriptscriptstyle +}}$ and in the interior of $M$).

\section{Concluding remarks}
\label{subsec:thermofield}

We discuss now a possible application of the eternal black hole Hamiltonian
(\ref{Hamiltonian})-(\ref{H+H-}) to the thermofield dynamics approach to
quantum
processes in black holes.
Thermofield dynamics is a theory of thermal fields based on the idea of
augmenting the
physical Fock space $\cal F$ by a fictitious Fock space  $\tilde {\cal F}$
obtained from the first one by a conjugation operation \cite{UmTa}. This
results
in a doubling of degrees of freedom for the system. A temperature dependent
``vacuum state" can be defined on the total Fock space  ${\cal F} \otimes
{\tilde {\cal F}}$ in such a way that vacuum expectation values of any physical
operator agree with their statistical average for an ensemble in thermal
equilibrium. A pure state  in  ${\cal F} \otimes {\tilde {\cal F}}$  will then
correspond to a mixed state in $\cal F$. This approach is self-consistent if
the total Hamiltonian for this extended system is of the general form
\begin{equation}
H = H_+ - H_-  \ , \label{tfdham}
\end{equation}
where $H_+$ is the Hamiltonian associated to $\cal F$ and   $H_-$ the
Hamiltonian
associated to $\tilde {\cal F}$.

In the original formulation of thermofield dynamics the space
$\tilde {\cal F}$ is merely formal. However, Israel \cite{Is} has suggested
that
a thermofield dynamics approach may give an adequate description of black hole
radiation and explain its thermal nature.
An `eternal' version of a black hole contains
two causally disconnected regions $R_+$ and $R_-$. The states of Boulware
particles in $R_+$ and $R_-$ can be identified with the Hilbert spaces
$\cal F$ and $\tilde {\cal F}$ respectively, so that the space
${\cal F} \otimes {\tilde {\cal F}}$ describes states of the complete system.
Israel's suggestion is the result of an explicit calculation that shows that an
observer in the right wedge $R_+$ observes particle modes travelling in the
background of a spacetime with a horizon as having thermal behavior \cite{Is}.
(For further development of this idea, see for example Refs.\cite{tfd}.)
This analysis is based only on the study of small perturbations
propagating in a black hole background. It is interesting to note that our
general result for the Hamiltonian for the
``tilted foliation" indicates that in the general case (even in one does not
restrict oneself by considering only small perturbations)
the complete Hamiltonian
(\ref{Hamiltonian}) of the system has the same structure as the Hamiltonian
(\ref{tfdham}), provided that the timelike surfaces bounding the system are
located
in different wedges. The Hamiltonians $H_+$ and $H_-$ can be used for the
definition of vacuum states and for the construction of two Fock spaces.
We have seen that the Hamiltonians $H_+$ and $H_-$
are encoded in the boundary conditions.  In a special case when the boundary
conditions on ${B_{\scriptscriptstyle +}}$ and ${B_{\scriptscriptstyle -}}$ are
the same, $H_+$ and $H_-$ are identical.
The property of the Hamiltonian mentioned above indicates that the thermofield
dynamics approach to black hole thermodynamics developed by Israel for small
perturbations might be naturally generalized to the situation when the
perturbations are not assumed to be small.

The description of dynamics of an eternal black hole requires data on two
timelike boundary surfaces $B_+$ and $B_-$.
It is natural to relate the data on $B_+$ and $B_-$  with
the external and internal degrees of freedom of a black hole respectively. This
opens the  possibility of generalizing the path integral approach to black hole
thermodynamics to the case when the internal degrees of freedom of the black
hole are present. It would be interesting to develop this analysis and apply it
to the problem of dynamical origin of black hole entropy.

\acknowledgments

The authors have benefitted from discussions with Andrei Barvinsky, Geoff
Hayward, and Andrei Zelnikov. They would also like to thank Werner Israel for
his interest and useful remarks.
Reseach support was received from  the Natural Sciences and Engineering
Research Council of Canada.



\vfill\eject
\centerline {\bf FIGURE CAPTIONS}
\smallskip
\parindent=0pt \parshape=3 0truein 6truein .7truein 5.3truein .7truein
5.3truein
Figure 1:~ Penrose diagram for a static, spherically symmetric spacetime. Each
point in
the diagram represents a topological sphere $S^2$. The bifurcation surface is
denoted by $S_0$. The spacetime is the union of four regions $R_\pm$ and
$T_\pm$ separated by the future and past horizons $H^\pm$.  $B_+$ and $B_-$
represent two timelike surfaces located in the regions $R_+$ and $R_-$
respectively.
The dashed lines in the figure represent  `initial' and `final' spacelike
hypersurfaces
${\Sigma}'$ and ${\Sigma}''$ of a ``tilted foliation". (For simplicity, tildes
have been omitted.)

\smallskip
\parindent=0pt \parshape=3 0truein 6truein .7truein 5.3truein .7truein
5.3truein
Figure 2:~ A slice $\Sigma$ (Einstein-Rosen bridge) of a static, spherically
symmetric eternal black hole. The surface $\Sigma$ has wormhole
topology  $R^1 \times S^2$. The parts of $\Sigma$ lying in $R_\pm$ are
${\Sigma}_{\pm}$ respectively, and $\Sigma ={\Sigma}_+ \cup {\Sigma}_-$.
The timelike boundaries $B_+$ and $B_-$ cross the Einstein-Rosen bridge at
two-dimensional surfaces $S_+$ and $S_-$. The bifurcation surface is denoted by
$S_0$. (Tildes have been omitted for simplicity.)

\smallskip
\parindent=0pt \parshape=3 0truein 6truein .7truein 5.3truein .7truein
5.3truein
Figure 3:~ A ``tilted foliation" $\Sigma$ of $M$ extending between the initial
(${\Sigma}'$)and final (${\Sigma}''$) spacelike hypersurfaces. Each point in
the diagram is
a topological sphere $S^2$. All the slices in the foliation intersect at the
bifurcation surface $S_0$.  The timelike boundaries $B_\pm$ intersect a
spacelike surface $\Sigma$ at the two disconnected two-dimensional surfaces
$S_\pm$.

\smallskip
\parindent=0pt \parshape=3 0truein 6truein .7truein 5.3truein .7truein
5.3truein
Figure 4:~ An ``untilted foliation" of $M$ extending between the hypersurfaces
${\Sigma}'$  and ${\Sigma}''$. Slices corresponding to different time
parameters do not intersect.

\smallskip
\parindent=0pt \parshape=3 0truein 6truein .7truein 5.3truein .7truein
5.3truein
Figure 5:~The quasilocal energy $E$ for a static, spherically symmetric
Einstein-Rosen bridge when the right boundary $S_+$ is  located at spatial
infinity. The location of the left boundary $S_-$ is denoted by $y_-$. The
horizon corresponds to $y_- =0$.

\vfill\eject
\end{document}